\documentclass[twocolumn,aps,showpacs,prb,tightenlines,amsmath,amssymb]{revtex4}

\usepackage{graphicx}
\usepackage{amssymb}
\usepackage{amsmath}
\usepackage{colordvi}
\newcommand{\bgreek}[1]{\mbox{\boldmath$#1$\unboldmath}}
\begin{document} 

\title{Spin diffusion in Si/SiGe quantum wells: spin relaxation in the
  absence of  D'yakonov-Perel' relaxation mechanism}

\author{P. Zhang}
\affiliation{Hefei National Laboratory for Physical Sciences at
  Microscale, University of Science and Technology of China, Hefei,
  Anhui, 230026, China}
\affiliation{Department of Physics,
  University of Science and Technology of China, Hefei,
  Anhui, 230026, China}
\author{M. W. Wu}
\thanks{Author to whom correspondence should be addressed}
\email{mwwu@ustc.edu.cn.}
\affiliation{Hefei National Laboratory for Physical Sciences at
  Microscale, University of Science and Technology of China, Hefei,
  Anhui, 230026, China}
\affiliation{Department of Physics,
  University of Science and Technology of China, Hefei,
  Anhui, 230026, China}
\altaffiliation{Mailing Address}

\date{\today}

\begin{abstract} 

In this work, the spin relaxation accompanying
the spin diffusion in symmetric
Si/SiGe quantum wells without the D'yakonov-Perel' 
spin-relaxation mechanism is calculated from a fully microscopic approach. The spin relaxation is caused by the
inhomogeneous broadening from the momentum-dependent
spin precessions in spatial domain under a magnetic field 
in the Voigt configuration. In fact, this inhomogeneous broadening
together with the scattering lead to an irreversible spin
relaxation along the spin diffusion. The effects of scattering,
magnetic field  and electron
density on spin diffusion are investigated. Unlike the case of spin
diffusion in the system with the D'yakonov-Perel' spin-orbit coupling such as
GaAs quantum wells where the
scattering can either enhance or reduce spin diffusion
depending on whether the system is in strong or weak scattering limit,
the scattering in the present system has no counter-effect on the inhomogeneous
broadening and suppresses the spin diffusion monotonically. 
The increase of magnetic
field reduces the spin diffusion,
while the increase of electron density enhances the spin diffusion when the
electrons are degenerate but has  marginal effect when the
electrons are nondegenerate. 

\end{abstract}

\maketitle

\section{Introduction}
The study of semiconductor spintronics has attracted a great
deal of attention for its potential application to spin-based
devices.\cite{aws,zutic,dy} Among different prerequisites for
realizing these devices, such as efficient spin
injection\cite{baidus,lombez} and suitable spin lifetime,\cite{nishikawa,kikkawa}
long spin diffusion/transport length\cite{kikkawa1,ivar,voros} is required sometimes,
especially for the design of spin transistor and spin
valve. Therefore, it is important to investigate the spin diffusion/transport in semiconductors.

In the study of spin diffusion/transport, the two-component
drift-diffusion model is widely used in the
literature.\cite{ivar,yu,yu1,fabian,huang,saikin,pershin} In this
model the spin-diffusion length $L_s$ is connected to spin-relaxation
time $\tau_s$ through spin-diffusion
coefficient $D_s$: $L_s=\sqrt{D_s\tau_s}$,\cite{fabian,ivar,yu,yu1} with
$D_s$ usually assumed to be equal to the charge-diffusion coefficient
$D_c$.\cite{yu,fabian,huang,pershin,saikin} This equation implies infinitely long
spin-diffusion length when the spin-relaxation time
$\tau_s$ goes to infinity. In bulk Si, there is
no D'yakonov-Perel' (DP) spin-orbit coupling\cite{dp} due to the bulk
inversion-symmetry. Thus the DP
spin-relaxation mechanism\cite{dp} is absent and the spin-relaxation
time in bulk Si is infinite when the other
spin-relaxation mechanisms (such as the Elliott-Yafet mechanism\cite{ey}) are
ignored. Therefore, an extremely long or even infinite spin-diffusion length is expected in bulk
Si. However, this is not the case in the presence of a magnetic field.
Very recently, Appelbaum {\sl et al.} studied the spin transport in
bulk Si with a magnetic field
perpendicular to both directions of spin transport and spin 
polarization.\cite{appelbaum} It has been shown
that the spin-diffusion length is very small when the magnetic field becomes slightly
strong (typically the spin-diffusion length is about several microns
when the magnetic field is in the order of 0.1~T).\cite{appelbaum} To account for the
experimental spin relaxation and dephasing (R\&D) along spin transport
as well as the small spin-diffusion length, the drift-diffusion model was 
utilized and the interference among different spin-precession angles
when reaching the same distance due to distinct transit times was
suggested to be important.\cite{huang} In fact, this spin
R\&D along spin diffusion due to the magnetic field in
the absence of the DP spin-relaxation mechanism was predicted from a
fully microscopic approach, i.e., the kinetic spin Bloch equation
(KSBE) approach,\cite{wu,wu-rev,clv,weng,weng1,cheng0,cheng} back in
2002.\cite{weng} In this approach, the momentum-dependent spin
precessions give rise to the inhomogeneous broadening.\cite{wu,wu-rev}
In the presence of the inhomogeneous broadening, any scattering
(including the spin-conserving scattering) leads to an irreversible spin
R\&D.\cite{wu,wu-rev,clv} In spin diffusion/transport, it
has been shown that the inhomogeneous broadening is determined by
the spin-precession frequency ${\bgreek \omega}_{\bf k}=\frac{m^\ast}
{\hbar^2 k_x}({\bf
  \Omega_k}+g\mu_B{\bf B})$ when the spin diffusion/transport is along
the $x$-axis.\cite{cheng} Here ${\bf \Omega_k}$ is the DP
term.\cite{dp} It was shown in Ref.~\onlinecite{weng} that even when
${\bf \Omega_k}=0$, the $k_x$ dependence in ${\bgreek \omega}_{\bf k}$
still causes spin R\&D. This is exactly the case 
in the experimental work on Si.
The present work is to investigate the spin diffusion in Si/SiGe
quantum wells (QWs) by means of the KSBE approach, 
in order to gain a deeper insight into
the spin relaxation along spin diffusion in the absence of the DP
spin-relaxation mechanism. The QWs are symmetric with even number of monoatomic Si layers and ideal
heterointerfaces to exclude the Rashba spin-orbit
coupling.\cite{golub} In addition, the study of spin
relaxation in asymmetric Si/SiGe QWs has been
carried out theoretically\cite{tahan} and experimentally,\cite{wil,jantsch}
showing that the Rashba spin-orbit coupling\cite{rashba} is very small (typically
about three orders of magnitude smaller than that in QW structures based on III-V semiconductors\cite{wil}) and
the spin-relaxation time is quite long (in the order of
$10^{-7}\sim10^{-5}$~s).\cite{tahan,wil,jantsch} Thus even for
asymmetric Si/SiGe QWs,
the present study still makes senses as long as the DP term ${\bf
  \Omega_k}$ is weak enough compared to the magnetic-field term $g\mu_B{\bf B}$.

\section{Model and KSBEs}
We start our investigation from an $n$-type symmetric Si/SiGe QW with
its growth direction along the $z||[001]$
direction. The lowest conduction band in bulk Si is located near the $X$ points
of the Brillouin zone. Due to the quantum confinement along the
$z$-direction, the two degenerate $X_z$ valleys lie lower than $X_x$
and $X_y$ valleys. The well width $a$ is set as small as 5
nm and the temperature is lower than 80 K, thus with the moderate electron concentration, only the lowest subband of $X_z$ valley is
occupied. The spin polarization is injected
constantly from one side of the sample
($x=0$ plane) with polarization $P$ and diffuses along the $x$-axis, while the
spatial distribution of spin in the $y$-direction is uniform. The magnetic
field ${\bf B}$ is applied in the $x$-$y$
plane. It can be along arbitrary direction without inducing any
essential difference in the absence of the DP spin-orbit
coupling. However, for the sake of convenience, the magnetic field is set to be
along the $y$-axis. 

As the two $X_z$ valleys are degenerate,
only one of them needs to be considered. However, both the
intra- and inter-valley scatterings have to be taken into account.
The KSBEs for one valley read\cite{weng,weng1,cheng0,cheng}
\begin{eqnarray}\nonumber
\frac{\partial \rho_{{\bf k}}(x,t)}{\partial t}&=&-\frac{e}{\hslash}\frac{\partial \Psi(x,t)}{\partial
  x}\frac{\partial \rho_{{\bf k}}(x,t)}{\partial
  k_{x}}-\frac{\hslash k_{x}}{m_t}\frac{\partial \rho_{{\bf k}}(x,t)}{\partial x}\\
\nonumber && -\frac{i}{\hslash}\left[\frac{g\mu_BB\sigma_y}{2}+\Sigma_{\bf k}(x,t),\rho_{{\bf
    k}}(x,t)\right]\\ &&+\left.\frac{\partial \rho_{{\bf
      k}}(x,t)}{\partial t}\right|_{scat}^{intra}+\left.\frac{\partial \rho_{{\bf
      k}}(x,t)}{\partial t}\right|_{scat}^{inter}.
\label{ksbe2}
\end{eqnarray}
Here $\rho_{\bf k}(x,t)$ represent the density matrices of electrons
with two-dimensional momentum ${\bf k}$ (referring to the bottom of $X_z$
valley under investigation) at position $x$ and time $t$. Their diagonal terms
$\rho_{{\bf k},\sigma\sigma}$$\equiv$$f_{{\bf
    k},\sigma}$ ($\sigma$=$\pm1/2$) represent the
electron-distribution functions and the off-diagonal ones $\rho_{{\bf
    k},1/2,-1/2}$=$\rho_{{\bf
    k},-1/2,1/2}^{\ast}$ describe the inter-spin-band correlations for the spin
 coherence. $\Sigma_{\bf k}(x,t)$ is the Hartree-Fock term from the
 Coulomb interaction.\cite{weng,weng1,cheng0} $-e$ is the electron
charge, $m_t=0.196 m_0$ is the transverse effective
mass in the $x$-$y$ plane, and $g\approx 2$ is the
effective $g$ factor for electrons in $X$ valleys of Si.\cite{graeff} $\Psi(x,t)$ is the electric
potential determined by the Poisson equation $\frac{\partial^2 \Psi(x,t)}{\partial
  x^2}=e[n(x,t)-N_0]/(4\pi\varepsilon_0\kappa_0 a)$ with
$n(x,t)=\sum_{\sigma}n_\sigma(x,t)$ standing for the electron density
at position $x$ and time $t$. $N_0$ is the background positive charge
density, and $n(x,0)=N_0$ denoting the initial
uniform spatial distribution of electron density. $\kappa_0=11.9$ is the relative
static dielectric constant.\cite{sze} $\left.\frac{\partial \rho_{{\bf 
      k}}(x,t)}{\partial t}\right|_{scat}^{intra}$ and $\left.\frac{\partial \rho_{{\bf
      k}}(x,t)}{\partial t}\right|_{scat}^{inter}$ originate from the
intra- and inter-valley scatterings, respectively. They are composed of the
electron-phonon, electron-impurity and electron-electron Coulomb
scatterings (see also the Appendix).

Before numerically solving the KSBEs, a much simplified situation with
the elastic electron-impurity scattering only is investigated
analytically. Based on this simplified investigation, some properties
of spin diffusion in Si/SiGe QWs can be speculated. Assuming ${\bf
  k}=k(\cos\theta,\sin\theta)$, in the steady state the Fourier
components of the density matrix with
respect to angle $\theta$ obey the following equation:
\begin{eqnarray}\nonumber
\frac{\hslash k}{2m_t}\frac{\partial}{\partial
  x}[\rho_k^{l+1}(x)+\rho_k^{l-1}(x)]&=&-i\frac{g\mu_B B}{2\hslash}[\sigma_y,\rho_k^l(x)]\\ 
&&-\frac{\rho_k^l(x)}{\tau_k^l},
\end{eqnarray}
with $\rho^l_k(x)=\frac{1}{2\pi}\int_0^{2\pi}d\theta\rho_{\bf
  k}(x)e^{-il\theta}$. Here $\frac{1}{\tau_k^l}=\frac{m^\ast
  N_i}{2\pi\hslash^3}\int_0^{2\pi}d\theta(1-\cos
l\theta)U_{\bf q}^2$ is the $l$th-order momentum-relaxation rate, with $|{\bf
  q}|=\sqrt{2k^2(1-\cos\theta)}$. $U_{\bf q}^2$ is the electron-impurity
scattering potential and $N_i$ represents the impurity density. It is noted that $\frac{1}{\tau_k^l}=\frac{1}{\tau_k^{-l}}$ and
$\frac{1}{\tau_k^0}=0$. From the above equation one can obtain a closed group of
first-order differential equations for $\rho_{k}^{\pm
  1}$ and $\rho_{k}^0$ by neglecting higher orders of $\rho_k^l$ with
$|l|>1$. From these equations the following second-order differential equation
about $\rho_k^0$ is obtained:
\begin{eqnarray}\nonumber
\frac{\partial^2}{\partial
  x^2}\rho_k^0(x)&=&-2\left(\frac{m_tg\mu_BB}{2\hslash^2
    k}\right)^2[\sigma_y,[\sigma_y,\rho_k^0(x)]]\\
&&+\frac{ig\mu_BB}{\hslash\tau_k^1}\left(\frac{m_t}{\hslash
    k}\right)^2[\sigma_y,\rho_k^0(x)].
\end{eqnarray}
Defining the ``spin vector'' as ${\bf S}_k^0(x)=\mbox{Tr}[\rho_k^0(x){\bgreek
  \sigma}]$ and using the boundary conditions (i) ${\bf
  S}_k^0(0)=(0,0,{S_k^0}_z)^{T}$ and (ii) ${\bf S}_k^0(+\infty)=0$,
${\bf S}_k^0(x)$ is found to be
\begin{eqnarray}\nonumber
{\bf
  S}_k^0(x)&=&{S_k^0}_z\left(\begin{array}{c}
    \sin\left[\frac{m_t}{\hslash k\tau_k^1}\left(\sqrt{1+\frac{1}{(\tau_k^1\omega_p)^2}}-1\right)^{-\frac{1}{2}}x\right] \\ 0 \\ 
    \cos\left[\frac{m_t}{\hslash
        k\tau_k^1}\left(\sqrt{1+\frac{1}{(\tau_k^1\omega_p)^2}}-1\right)^{-\frac{1}{2}}x\right]\end{array}\right)\\ &&
\times e^{-\frac{m_t\omega_p}{\hslash
    k}\left(\sqrt{1+\frac{1}{(\tau_k^1\omega_p)^2}}-1\right)^{\frac{1}{2}}x},
\label{eq2}
\end{eqnarray}
with $\omega_p=g\mu_BB/\hslash$ being the spin-precession frequency in
time domain under magnetic field ${\bf B}=B\hat y$. The total spin
signal in the $z$-direction (including two valleys) reads $S_z(x)=2\sum_{\bf
  k}\mbox{Tr}[\rho_{\bf k}(x)\sigma_z]=\int_{0}^{+\infty}\frac{dk}{\pi}k{S_k^0}_z(x)$. To get
$S_z(x)$, ${S_k^0}_z$ in boundary condition (i) is
assumed to be ${S_k^0}_z=f_{k,1/2}-f_{k,-1/2}$ with
$f_{k,\sigma}=\frac{1}{e^{\beta(\hslash^2k^2/2m_t-\mu_{\sigma})}+1}$.
Here $\mu_\sigma$ is determined by
$\int_{0}^{+\infty}\frac{dk}{\pi}k(f_{k,1/2}+f_{k,-1/2})=N_0$ and
$\int_{0}^{+\infty}\frac{dk}{\pi}k(f_{k,1/2}-f_{k,-1/2})=PN_0$ with
$P$ the spin polarization at $x=0$ plane.

It is noted from Eq.~(\ref{eq2}) that during the spin diffusion,
${\bf S}_k^0(x)$ on one hand precesses around the direction
of magnetic field, and on the other hand decays in
magnitude. However, when $\tau_k^1$ becomes infinity, i.e., in the
limit of zero electron-impurity scattering, Eq.~(\ref{eq2})
becomes ${\bf S}_k^0(x)={S_k^0}_z\left(\sin\frac{\sqrt{2}m_t\omega_p}{\hbar k}x,0,\cos\frac{\sqrt{2}m_t\omega_p}{\hbar k}x\right)^T$.
This solution clearly indicates the momentum dependence
of spin precession in spatial domain under magnetic field, which leads to
the inhomogeneous broadening.\cite{weng} This inhomogeneous broadening alone
leads to a reversible decay of the total spin signal $S_z(x)$
along the $x$-direction. However, in the presence of
scattering, this decay becomes irreversible, as shown in the
exponential damping term for each $S_{kz}^0(x)$ in
Eq.~(\ref{eq2}). Therefore, even without the DP term, there is spin
relaxation accompanying the spin diffusion. Due to the factor
$S_{kz}^0$ included in Eq.~(\ref{eq2}),
the main  contribution to $S_z(x)$ comes from $k$-states near the 
Fermi surface at $x=0$ plane, especially when the
temperature is low. Thus 
\begin{eqnarray}\nonumber
S_{z}(x)\propto&&\cos\left[\frac{m_t}{\hslash
    k_f\tau_{k_f}^1}\left(\sqrt{1+\frac{1}{(\tau_{k_f}^1\omega_p)^2}}
-1\right)^{-\frac{1}{2}}x\right]\\
  &&\times e^{-\frac{m_t\omega_p}{\hslash k_f}\left(\sqrt{1+\frac{1}
{(\tau_{k_f}^1\omega_p)^2}}-1\right)^{\frac{1}{2}}x}\ ,
\label{eq3}
\end{eqnarray}
indicating that the spin diffusion can be suppressed by the
scattering strength and the magnetic field, based on the increase of
the exponential damping rate with $1/\tau_{k_f}^1$
and $\omega_p$. It has been shown earlier
that, in systems with the DP spin-orbit coupling, the scattering on one
hand provides the spin-R\&D channel in the presence of the inhomogeneous
broadening from the DP term but, on the other
hand, can have a counter-effect on the inhomogeneous
broadening.\cite{wu-rev,clv,cheng} In the strong scattering limit, the counter-effect dominates and thus increasing
scattering strength results in a suppression of spin R\&D
in time domain\cite{wu-rev,clv} or an enhancement of spin diffusion/transport in spatial
domain.\cite{cheng} However, for the current case without the DP spin-orbit coupling, the scattering shows
{\em no} counter-effect on the inhomogeneous broadening but just suppresses the spin
diffusion. Except for the above analysis based on the
simplified model, to gain a complete picture of the problem
of spin diffusion, one 
must solve the KSBEs taking into account the electron-phonon and
electron-electron Coulomb scatterings, both of which play an important role 
on spin R\&D. 

\section{Numerical results}
\begin{figure}[ht]
    {\includegraphics[width=8.8cm]{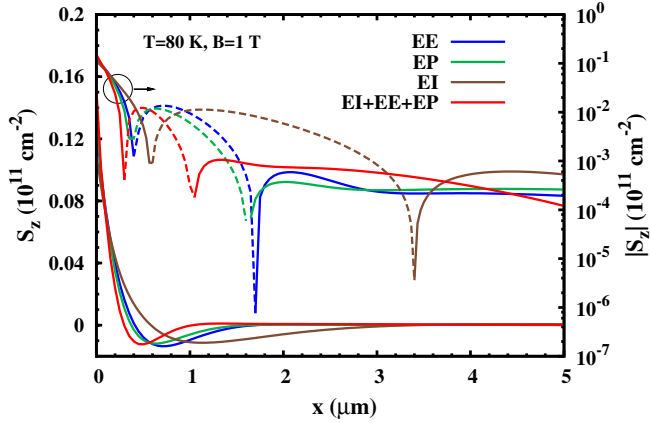}}
    \caption{(Color online) The steady-state spatial distributions of spin
      signal $S_z$ with different scatterings included. The
      curve labeled  ``EE'', ``EP'' or ``EI'' stands
 for the calculations with  the electron-electron,
 electron-phonon or electron-impurity scattering, respectively,
      while the curve labeled 
``EI+EE+EP'' stands for the calculation with all the
 scatterings. In order to get a clear view of the decay
 and precession of $S_z$, we also plot the
corresponding absolute value of $S_z$ against $x$ 
on a log-scale (Note the scale is on the right hand of the frame).
The dashed curves correspond to the part with  
$S_z<0$.
}
  \label{fig1}
\end{figure}

\begin{figure}[ht]
    {\includegraphics[width=9.7cm]{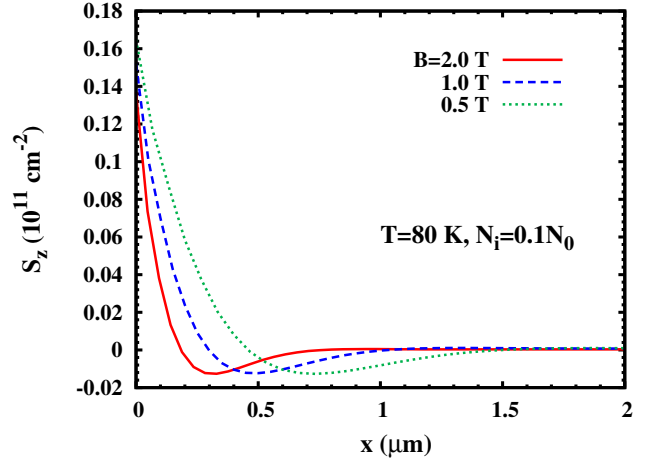}}
    \caption{(Color online) $S_z$ {\em vs.} $x$ in the steady state
for different magnetic field strengths. 
Solid curve: $B=2$~T; Dashed curve: $B=1$~T;
 Dotted curve: $B=0.5$~T. $T=80$~K and $N_{i}=0.1N_0$.}
  \label{fig2}
\end{figure}

\begin{figure}[ht]
    {\includegraphics[width=9.9cm]{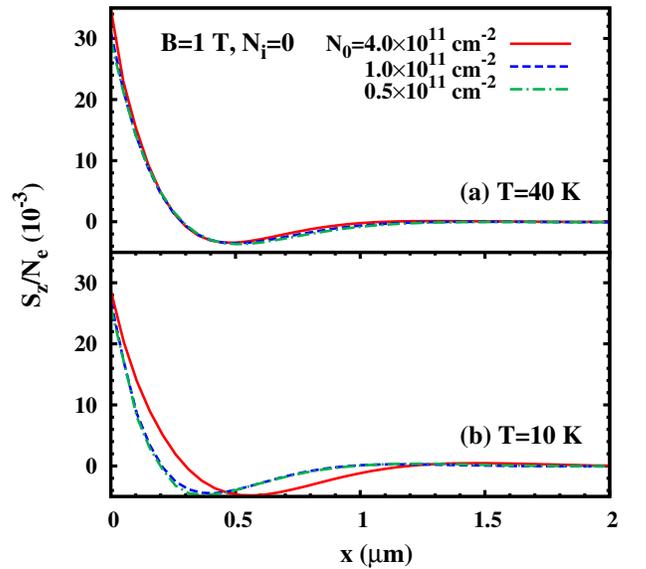}}
    \caption{(Color online) $S_z/N_0$ {\em vs.} $x$ in the steady state
with different electron densities at (a) $T=40$~K and (b) $T=10$~K. 
 $B=1$~T and $N_i=0$.}
  \label{fig3}
\end{figure}

To numerically solve the KSBEs, the double-side boundary conditions
are used.\cite{cheng} These conditions assume a
 steady spin polarization $P$ for electrons
with $k_x>0$ at boundary $x=0$ and a vanishing spin polarization at
finite sample length $x=L$ for electrons with $k_x<0$.\cite{cheng} For the Poisson
equation, the boundary conditions
are set to be $\Psi(0,t)=\Psi(L,t)=0$. The numerical scheme for solving the
KSBEs is given in Ref.~\onlinecite{cheng}. Once the
KSBEs are numerically solved, the steady-state distribution of electrons
$N_\sigma(x)=n_\sigma(x,+\infty)=2\sum_{\bf k}f_{{\bf k},\sigma}(x,+\infty)$
(factor 2 comes from the two degenerate $X_z$ valleys) is calculated 
and thus the spin signal
$S_z(x)=N_{1/2}(x)-N_{-1/2}(x)$ can be derived to analyze the
spin-diffusion properties. In the calculation, the electron
density $N_0$ is set to be 4$\times 10^{11}$~cm$^{-2}$ 
except otherwise specified, and the impurity
density $N_i$ is assumed to be 0.1$N_0$ when the
impurities are present. Furthermore, the initial spin
polarization $P$ at $x=0$ plane is set to be 5~\%. The effects
of scattering, magnetic field and electron density on spin diffusion are investigated, with the main results 
given in Figs.~\ref{fig1}-\ref{fig3}.

We first investigate the effects of scattering and magnetic field on
spin diffusion.
The steady-state spatial distributions of spin signal calculated with different
scattering are shown in
Fig.~\ref{fig1}. In order to show the property of
spin diffusion clearly, we also plot the absolute value of $S_z$ 
{\em vs.} $x$ on a log-scale in the same figure.
All these curves indicate obvious spin relaxation
along spin diffusion without the DP spin-relaxation
mechanism. By comparing the curves labeled as ``EE'' and ``EI'',
one finds that the electron-electron Coulomb scattering can
suppress spin diffusion effectively. In fact, the Coulomb
scattering plays an important role in both spin
R\&D\cite{wu-rev,clv,glazov} and spin
diffusion/transport.\cite{weng1,cheng0,cheng} In a system with the DP spin-orbit
coupling, the Coulomb scattering 
not only contributes to the total momentum relaxation time
$\tau_k$,\cite{glazov} but also has a counter-effect to the inhomogeneous
broadening.\cite{wu-rev,clv,weng1,cheng0,cheng} Therefore, in the
strong scattering limit, adding Coulomb scattering may suppress the
spin relaxation and enhance the spin diffusion/transport.\cite{clv,cheng0}
For the current
situation without the DP spin-orbit coupling, the Coulomb scattering
affects the spin diffusion only through $\tau_k$ and thus only
suppresses the spin diffusion/transport. Similarly, the
electron-phonon scatterings, including both the intra- and
inter-valley scatterings, also contribute to the momentum
relaxation and suppress spin diffusion effectively, as shown by the
``EP'' curve. For the case of stronger scattering strength
(i.e., with all the different scatterings included), the
spin-diffusion length becomes much smaller. It is also noted that with
the increase of scattering strength, the spatial spin-precession period
decreases. This is because the spin-precession frequency increases with
$1/\tau_{k_f}^1$ as shown in Eq.~(\ref{eq3}). The magnetic-field
dependence of spin diffusion is investigated as well, with the results
corresponding to different magnetic fields shown in
Fig.~\ref{fig2}. It is revealed that both the spin-diffusion length and the spin-precession
period decrease with an increase of
magnetic field strength $B$. This is because both the damping rate and the spin-precession frequency increase with
$\omega_p$ and thus $B$, as shown in Eq.~(\ref{eq3}). These studies
indicate a suppression of spin
diffusion due to the scattering and
magnetic field, just as obtained earlier with the simplified model.


The density dependence of spin diffusion is also investigated. The
steady-state spatial distributions of spin signal with three
different electron densities under temperature $T=40$ and 10~K are
plotted in Fig.~\ref{fig3}(a) and (b), respectively. The spin signal is
rescaled by the corresponding electron density to be $S_z/N_0$ for comparison. It is noted that when
$T=40$~K, the spin diffusion is {\em insensitive} to the electron
density. That is because the electrons are nondegenerate in the
studied electron-density regime when $T=40$~K, due to
the large transverse effective electron mass in $X$ valleys of Si. However, when $T$ decreases
to 10~K, the effect of electron density on the spin diffusion
can be seen, as shown in Fig.~\ref{fig3}(b). 
In the large density regime where the electrons become degenerate, the spin
diffusion length increases with the electron density (compare the situations with $N_0=4.0\times 10^{11}$ and 1.0$\times 10^{11}$~cm$^{-2}$). This is mainly due to the
decrease of the damping rate with $k_f$, as shown in
Eq.~(\ref{eq3}). In addition, in the low density regime it is shown
that the density again has a marginal effect
on spin diffusion (compare the situations with $N_0=1.0\times
10^{11}$ and 0.5$\times 10^{11}$~cm$^{-2}$), as the electrons remain
nondegenerate there. It is noted that the density dependence 
of spin diffusion in Si is very different from the density
dependence of the spin relaxation in GaAs QWs where non-monotonic
density dependence was predicted\cite{clv} and 
realized experimentally very recently.\cite{teng} 
This is due to the fact that in GaAs QWs the inhomogeneous broadening comes
from the DP term which is, however, absent in the symmetric Si/SiGe
QWs.

\section{Summary}
In summary, the present work investigates the spin diffusion in
symmetric Si/SiGe (001) QWs at low temperature. There is no DP
spin-relaxation mechanism due to the absence of the DP spin-orbit coupling in
this system. However, a magnetic field in the Voigt configuration is
present. Our simulations were performed in a
fully microscopic way based on the KSBE approach, with
all the relevant scatterings included. It was shown that, even without
the DP spin-relaxation
mechanism, the electron spins relax effectively along 
the spin diffusion. This spin relaxation is
caused by the inhomogeneous broadening from the momentum-dependent
spin precessions in spatial domain. The effects of scattering, magnetic
field and electron density on spin diffusion were
investigated. 
It was shown that, unlike the case of spin diffusion in the system with
the DP spin-orbit coupling,\cite{cheng} in Si/SiGe (001) QWs any
scattering suppresses the spin diffusion without any counter-effect on the
inhomogeneous broadening. The magnetic field reduces spin diffusion
also. It was further revealed that the 
increase of electron density enhances the spin diffusion
when the electrons are degenerate but has  marginal effect when the
electrons are nondegenerate. 

\begin{acknowledgments}
This work was supported by the Natural Science Foundation of China
under Grant No.~10725417, the
National Basic Research Program of China under Grant 
No.~2006CB922005 and the Knowledge Innovation Project of Chinese Academy
of Sciences. 
\end{acknowledgments}

\begin{appendix}
\section{The scattering terms of the KSBEs}
The scattering terms are analogous to those shown in 
Ref.~\onlinecite{zhang}, except the following differences. The two valleys
are degenerate here, thus the multivalley KSBEs similar to those shown
in Ref.~\onlinecite{zhang} can be simplified to obtain the KSBEs of a ``single''
valley [i.e., Eq.~(\ref{ksbe2})]. 
The intra-valley scatterings consist of those due to 
longitudinal-acoustic (LA) and transverse-acoustic (TA)
phonons, while the inter-valley scatterings are the $g$-type
scatterings involving LA, TA and longitudinal-optical (LO)
 phonon branches.\cite{pop} However, due to the low temperature, the
 scattering due to the LO-phonon is
 neglected. $M^2_{\alpha,intra,{\bf Q}}=
 \frac{\hslash
   D_{\alpha}^2Q^2}{2d\Omega_{\alpha,intra,{\bf
       Q}}}|I_{intra}(iq_{z})|^{2}$ is the matrix element for the intra-valley scattering, and $M^2_{\alpha,inter,{\bf Q}}=
 \frac{\hslash
   \Delta_{\alpha}^2}{2d\Omega_{\alpha,inter,{\bf
       Q}}}|I_{inter}(iq_{z})|^{2}$ for the inter-valley
 scattering. $\alpha=LA/TA$ stands for the LA/TA phonon mode. $d=2.33$~g/cm$^3$ is the mass density of Si.\cite{sonder} $D_{LA}=6.39$~eV and
 $D_{TA}=3.01$~eV.\cite{pop} $\Omega_{\alpha,intra,{\bf Q}}=v_\alpha Q$
 with phonon velocities $v_{LA}=9.01\times 10^5$~cm/s and
 $v_{TA}=5.23\times 10^5$~cm/s.\cite{pop} $\Delta_{LA}=1.5\times 10^8$~eV/cm and $\Delta_{TA}=0.3\times 10^8$
 eV/cm.\cite{pop} The phonon energies for the inter-valley scattering are
 approximately fixed to be $\hslash\Omega_{LA,inter,{\bf Q}}=0.019$~eV and
 $\hslash\Omega_{TA,inter,{\bf Q}}=0.01$~eV.\cite{pop} $|I_\gamma(iq_z)|^2=
 \frac{\pi^{4}\sin^{2}y}{y^{2}(y^{2}-\pi^{2})^{2}}$
 ($\gamma=intra/inter$) is the form factor
 with $y\equiv aq_{z}/2$ for the intra-valley scattering and $y\equiv
 a(q_{z}-2K_{X_z}^z)/2$ for the inter-valley scattering. Here
 $K_{X_z}^z=0.85\times\frac{2\pi}{a_0}$ with $a_0$ being the Si
 lattice constant is the
 $z$ component of the coordinate of the bottom of $X_z$ valley.
\end{appendix}


\begin{thebibliography}{0}
\bibitem{aws}D. D. Awschalom, D. Loss, and N. Samarth, {\it
    Semiconductor Spintronics and Quantum Computation} (Springer,
  Berlin, 2002).
\bibitem{zutic}I. \v Zuti\' c, J. Fabian, and S. D. Sarma,
  Rev. Mod. Phys. {\bf 76}, 323 (2004);  J. Fabian, A. Matos-Abiague, C. Ertler,
P. Stano, and I. \v Zuti\'c, Acta Phys. Slovaca {\bf 57}, 565
(2007).
\bibitem{dy}M. I. D'yakonov, {\it Spin Physics in Semiconductors}
  (Springer, Berlin, 2008).

\bibitem{baidus}N. V. Baidus, M. I. Vasilevskiy, M. J. M. Gomes,
  M. V. Dorokhin, P. B. Demina, E. A. Uskova, B. N. Zvonkov,
  V. D. Kulakovskii, A. S. Brichkin, A. V. Chernenko, and
  S. V. Zaitsev, Appl. Phys. Lett. {\bf 89}, 181118 (2006).
\bibitem{lombez}L. Lombez, P. Renucci, P. F. Braun, H. Carr\`ere,
  X. Marie, T. Amand, B. Urbaszek, J. L. Gauffier, P. Gallo, T. Camps,
  A. Arnoult, C. Fontaine, C. Deranlot, R. Mattana, H. Jaffr\`es,
  J. -M. George, and P. H. Binh, Appl. Phys. Lett. {\bf 90}, 081111
  (2007).

\bibitem{nishikawa}Y. Nishikawa, A. Takeuchi, M. Yamaguchi, S. Muto,
  and O. Wada, IEEE J. Quantum Electron {\bf 2}, 661 (1996).
\bibitem{kikkawa}J. M. Kikkawa, I. P. Smorchkova, N. Samarth, and
  D. D. Awschalom, Science {\bf 277}, 1284 (1997).
\bibitem{kikkawa1}J. M. Kikkawa and D. D. Awschalom, Nature (London) {\bf
    397}, 139 (1999).
\bibitem{ivar}I. Martin, Phys. Rev. B {\bf 67}, 014421 (2003).
\bibitem{voros}Z. V\"or\"os, R. Balili, D. W. Snoke, L. Pfeiffer, and
  K. West, Phys. Rev. Lett. {\bf 94}, 226401 (2005).
\bibitem{yu}Z. G. Yu and M. E. Flatt\'e, Phys. Rev. B {\bf 66},
  201202 (2002).
\bibitem{yu1}Z. G. Yu and M. E. Flatt\'e, Phys. Rev. B {\bf 66}, 235302 (2002).
\bibitem{fabian}J. Fabian, I. \v Zuti\'c, and S. D. Sarma, Phys. Rev. B
  {\bf 66}, 165301 (2002).

\bibitem{huang}B. Huang and I. Appelbaum, Phys. Rev. B {\bf 77},
  165331 (2008).
\bibitem{saikin}S. Saikin, J. Phys.: Condens. Matter {\bf 16}, 5071
  (2004).
\bibitem{pershin}Y. V. Pershin, Physica E {\bf 23}, 226 (2004).


\bibitem{dp}M. I. D'yakonov and V. I. Perel',
  Zh. \'Eksp. Teor. Fiz. {\bf 60}, 1954 (1971) [Sov. Phys. JETP {\bf
    33}, 1053 (1971)].
\bibitem{ey}R. J. Elliott, Phys. Rev. {\bf 96}, 266 (1954). 

\bibitem{appelbaum}I. Appelbaum, B. Huang, and D. J. Monsma, Nature
  {\bf 447}, 295 (2007).


\bibitem{wu}M. W. Wu and C. Z. Ning, Eur. Phys. J. B {\bf 18}, 373
  (2000); M. W. Wu, J. Phys. Soc. Jpn. {\bf 70}, 2195 (2001).
\bibitem{wu-rev}M. W. Wu, M. Q. Weng, and J. L. Cheng,
in {\it Physics, Chemistry and
  Application of Nanostructures: Reviews and Short Notes to
  Nanomeeting 2007}, edited by V. E. Borisenko, V. S. Gurin, and
S. V. Gaponenko (World Scientific, Singapore, 2007), p. 14, and references
therein.
\bibitem{clv}C. L\"u, J. L. Cheng, and M. W. Wu, Phys. Rev. B {\bf
    73}, 125314 (2006).

\bibitem{weng}M. Q. Weng and M. W. Wu, Phys. Rev. B {\bf 66}, 235109
  (2002).
\bibitem{weng1}M. Q. Weng and M. W. Wu, J. Appl. Phys. {\bf 93}, 410
(2003).
\bibitem{cheng0}J. L. Cheng, M. W. Wu, and I. C. da Cunha Lima, Phys. Rev. B
{\bf 75}, 205328 (2007).

\bibitem{cheng}J. L. Cheng and M. W. Wu, J. Appl. Phys. {\bf 101},
073702 (2007).

\bibitem{golub}L. E. Golub and E. L. Ivchenko, Phys. Rev. B {\bf 69},
  115333 (2004).
\bibitem{tahan}C. Tahan and R. Joynt, Phy. Rev. B {\bf 71}, 075315
  (2005).
\bibitem{wil}Z. Wilamowski, W. Jantsch, H. Malissa, and U. R\"ossler,
  Phys. Rev. B {\bf 66}, 195315 (2002).
\bibitem{jantsch}W. Jantsch, Z. Wilamowski, N. Sandersfeld,
  M. M\"uhlberger, and F. Sch\"affler, Physica E {\bf 13}, 504
  (2002). 

\bibitem{rashba}Y. Bychkov and E. Rashba, J. Phys. C {\bf 17}, 6039 (1984). 

\bibitem{graeff}C. F. O. Graeff, M. S. Brandt, M. Stutzmann,
  M. Holzmann, G. Abstreiter, and F. Sch\"affler, Phys. Rev. B {\bf
    59}, 13242 (1999).
\bibitem{sze}S. M. Sze, {\sl Physics of Semiconductor Devices} (Wiley-Interscience, New
York, 1981), p. 849.

\bibitem{glazov}M. M. Glazov and E. L. Ivchenko,
  Zh. Eksp. Teor. Fiz. {\bf 126}, 1465 (2004) [JETP {\bf 99}, 1279 (2004)].
\bibitem{teng} L. H. Teng, P. Zhang, T. S. Lai, and M. W. Wu, 
Europhys. Lett. {\bf 84}, 27006 (2008). 
\bibitem{zhang}P. Zhang, J. Zhou, and M. W. Wu, Phys. Rev. B {\bf 77},
  235323 (2008).

\bibitem{pop}E. Pop, R. W. Dutton, and K. E. Goodson,
  J. Appl. Phys. {\bf 96}, 4998 (2004).
\bibitem{sonder}E. Sonder and D. K. Stevens, Phys. Rev. {\bf 110}, 1027 (1958).


\end{thebibliography}
\end{document}